\documentclass[letter]{aa}
\usepackage{graphicx}
\usepackage{braket}
\usepackage{amsmath}
\usepackage{mathtools}
\usepackage{multicol}
\usepackage{xcolor}
\usepackage{ulem}
\usepackage{caption}
\DeclarePairedDelimiterX{\norm}[1]{\lVert}{\rVert}{#1}

\newcommand{\dt}[1]{\frac{\partial#1}{\partial t}}
\newcommand{\dr}[1]{\frac{\partial#1}{\partial r}}

\newcommand{\dm}[1]{\frac{\partial#1}{\partial m}}
\newcommand{\curl}{\Vec{\nabla}\wedge}

\newcommand{\numax}{\nu_{\text{max}}}
\newcommand{\uHz}{\mu\text{Hz}}
\newcommand{\Msun}{\text{M}_\odot}
\newcommand{\tcool}{t_\text{cool}}
\newcommand{\rthresh}{r_\text{thresh}}

\usepackage{natbib,twoopt}
\usepackage[breaklinks=true]{hyperref}
\bibpunct{(}{)}{;}{a}{}{,} 
\makeatletter
  \newcommandtwoopt{\citeads}[3][][]{\href{http://adsabs.harvard.edu/abs/#3}
    {\def\hyper@linkstart##1##2{}
     \let\hyper@linkend\@empty\citealp[#1][#2]{#3}}}
  \newcommandtwoopt{\citepads}[3][][]{\href{http://adsabs.harvard.edu/abs/#3}
    {\def\hyper@linkstart##1##2{}
     \let\hyper@linkend\@empty\citep[#1][#2]{#3}}}
  \newcommandtwoopt{\citetads}[3][][]{\href{http://adsabs.harvard.edu/abs/#3}
    {\def\hyper@linkstart##1##2{}
     \let\hyper@linkend\@empty\citet[#1][#2]{#3}}}
  \newcommandtwoopt{\citeyearads}[3][][]
    {\href{http://adsabs.harvard.edu/abs/#3}
    {\def\hyper@linkstart##1##2{}
     \let\hyper@linkend\@empty\citeyear[#1][#2]{#3}}}
\makeatother

\newcommand\orc[1]{\href{https://orcid.org/#1}{\includegraphics[width=3mm]{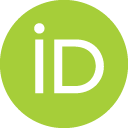}}}

\begin{document}
\title{Magneto-Archeology of White Dwarfs}
\subtitle{Revisiting the fossil field scenario with observational constraints during the red giant branch}
\titlerunning{Magneto-Archeology of White Dwarfs}
\authorrunning{Einramhof et al.}
\author{
    L. Einramhof\inst{1}\orc{0009-0002-5619-0598},
    {L. Bugnet}\inst{1}\orc{0000-0003-0142-4000},
    {L. M. Calcaferro}\inst{2,3}\orc{0000-0002-1345-8075},
    {L. Barrault}\inst{1}\orc{},
    and
    {S. B. Das}\inst{4}\orc{0000-0003-0896-7972}
}

\institute{
   Institute of Science and Technology Austria (ISTA), Am Campus 1, Klosterneuburg, Austria\\
    \email{lukas.einramhof@ista.ac.at} 
    \and
    Grupo de Evolución Estelar y Pulsaciones, Facultad de Ciencias Astronómicas y Geofísicas, Universidad Nacional de La Plata, Paseo del Bosque s/n, (1900) La Plata, Argentina
    \and
    Instituto de Astrofísica La Plata, CONICET-UNLP, Paseo  del Bosque s/n, (1900) La Plata, Argentina
    \and
    Center for Astrophysics | Harvard \& Smithsonian, 60 Garden Street, Cambridge, MA 02138, USA\\
}

\date{Received 21.01.2026; accepted 16.03.2026}

\abstract{
The detection of strong, large-scale magnetic fields at the surface of mainly the oldest population of white dwarfs might point towards a hidden internal magnetic field slowly rising to the surface. In addition, strong magnetic fields have recently been measured through asteroseismology in the radiative interiors of red giant stars, the progenitors of white dwarfs.
To investigate the potential connection between these observations, we revisit the fossil field framework by using the asteroseismic detections to constrain the strength of such magnetic fields as they evolve to the white dwarf stage. We assume that the magnetic field was either created during the core convection on the main sequence or that it fills the radiative interior as the star evolves on the red giant branch. From these, we evolve the magnetic flux, allowing for magnetic diffusion along the evolution of a 1.5$\Msun$ modelled star.
We find that measured field strengths in red giants attributed to the hydrogen-burning shell are compatible with the field amplitudes and emergence timescales of magnetized white dwarfs. 
On the contrary, magnetic fields generated solely from a convective-core dynamo on the main-sequence and detectable during the red giant branch would be buried too deep in the star and not match the breakout timescales and the field strengths of magnetic white dwarfs. A broadly magnetized internal radiative zone during the red giant branch is therefore key for the fossil field theory to connect magnetic fields observed along the late evolution of stars.
}


\maketitle
\nolinenumbers

\section{Introduction}
White dwarfs (WDs) give us key insights into the properties of the cores of their progenitors. In particular, exposed stable magnetic fields at the surface of WDs might be related to past core magnetism \citep[e.g.][]{Quentin2018, Ferrario2020}, a crucial player in determining the internal distribution of angular momentum along the evolution of stars \citep[e.g.][]{Takahashi2021}.

\begin{figure*}[ht]
    \centering
    \includegraphics[width=1\textwidth]{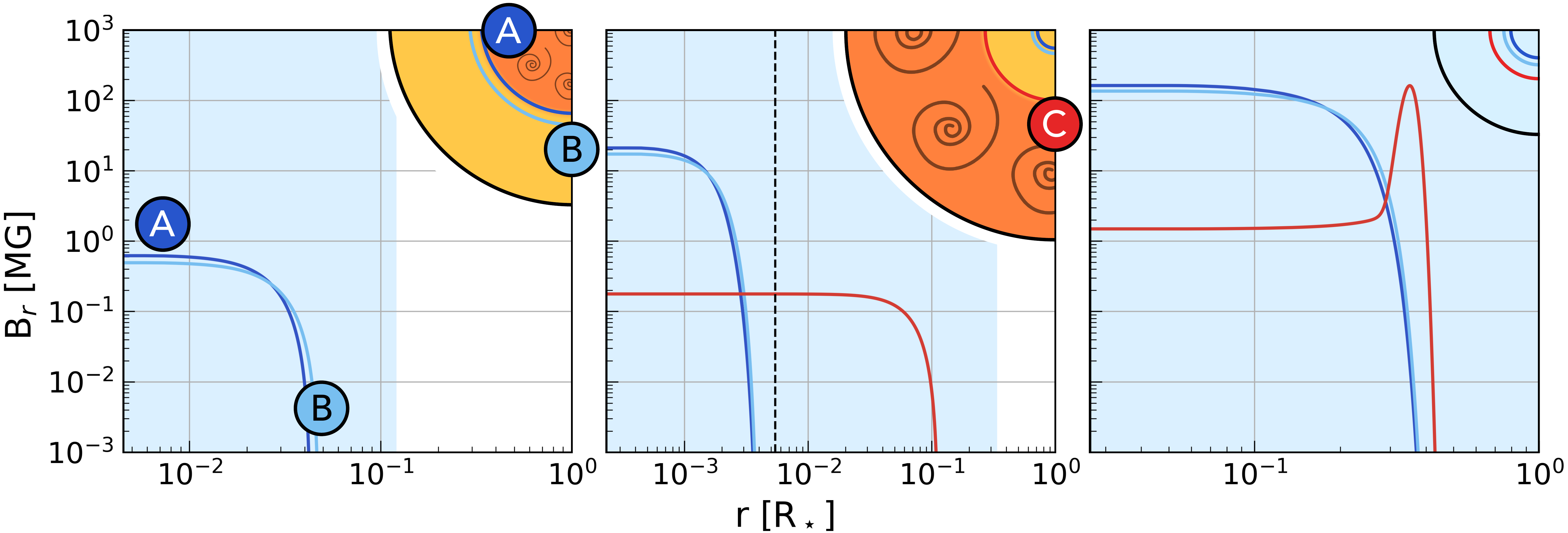}
    \caption{Considered magnetic field configurations at different evolutionary stages of a typical $1.5\Msun$ star. The central field strengths are set such that asteroseismic detections of all three fields would measure  100kG during the RGB (see App.~\ref{app:kernel_avg}). Left panel: Scenarios A (darkblue line) and B (lightblue line) are created by the convective core during the MS and start evolving as a large-scale stable field at the end of the MS. Middle panel: Magnetic fields during the RGB when the typical oscillation frequency of the star reaches $150\uHz$. The two fields for scenarios A and B are now buried below the Hydrogen-burning shell (dashed line). Scenario C (red line) starts its evolution here and fills the entire radiative interior. Right panel: Magnetic field configurations for all three scenarios at the start of the WD cooling sequence. The blue shaded region corresponds to the radial extent of the WD mass at different evolutionary stages.}
    \label{fig:intro}
\end{figure*}

\citet{Bagnulo2022} and \citet{Moss2025} measured the surface magnetic fields of close-by WDs. They find an incidence of surface magnetism of roughly 6\% for WDs with field strengths of $\sim1-100$MG. This increases to 20\% if also weakly magnetized WDs ($<1$MG) are considered \citep{Bagnulo2022}.
In addition, the incidence of magnetism increases with age for typical WDs resulting from single-star evolution, in the mass range 0.5-1$\Msun$. Two key questions are therefore where these surface stable magnetic fields originate from, and why their emergence seems to be delayed along the cooling sequence. Excluding binary interaction channels, magnetic fields at the surface of WDs might likely be created either by a crystallization dynamo \citep[e.g.][]{Isern2017, Ginzburg2022, Fuentes2024, Castro-Tapia2024} or via the fossil field scenario (hereafter FFS), which was for instance studied recently by \citet[][hereafter C24]{Camisassa2024} and \citet[][hereafter CT26]{Castro-Tapia2026}.
C24 and CT26 demonstrate that, using typical dynamo efficiency in the convective core during the main sequence (MS) and conserving the magnetic flux all the way to the WD phase, the FFS leads to an emergence of magnetism at the surface of WDs in timescales compatible with observations. However, one key additional observational constraint absent from the picture is the recent detection of magnetic fields in the interior of WD progenitors, red giants (RGs). Magnetic fields have been detected in the radiative interiors of RGs with a sample of a few tens confirmed magnetic RG cores to date \citep[][and references within]{Hatt2024}.
We revisit the FFS, including these new asteroseismic constraints from early stages, and investigate when and with what properties resulting fossil fields could explain the emergence of magnetism at the surface of old WDs.

\section{Magnetic evolution framework}
\subsection{Magnetic evolution scenarios and configurations}
We consider three different origins and initial setups for the evolution of the magnetic field. First, we assume that the field comes from the convective core during the MS and starts the evolution at the end of the MS (Scenario A, Fig.~\ref{fig:intro} dark blue). Similarly, Scenario B assumes the same but also allows for flexibility on the magnetic extent and for the diffusion of the magnetic field into the radiative envelope during the MS, resulting in a larger magnetic region at the end of the MS (Scenario B, Fig.~\ref{fig:intro} light blue). Lastly, we place a stable field directly in the full radiative interior during the RGB (Scenario C, Fig.~\ref{fig:intro} red). Possible formation channels for such a field and additional details on all three scenarios are given in App.~\ref{app:mag_scenarios}. 
The initial internal fields' extent is limited to the radiative zone, and we neglect the impact of convection and convective dynamo action during red giant stages on the diffusion of the large-scale internal magnetic field, due to the expected slow motion and low efficiency of the dynamo action \citep{Auriere2015, Amard2024}, and unconstrained coupling with the stable field.
We consider dipolar magnetic fields extracted from the most stable initial field configurations, following the works of \cite{Broderick2007, Braithwaite2008, Duez2010}. Additional information on the magnetic field profiles can be found in App.~\ref{app:mag_geom}.

\subsection{Observational constraints from Asteroseismology}
To calibrate the strength of the fields we consider, we use constraints from the largest sample of magnetic measurements in the radiative interior of RGs \citep{Hatt2024}.
We evolve a representative model from their sample with a mass of $1.5\Msun$ (Z=0.02) using Modules for Experiments in Stellar Astrophysics \citep[\texttt{MESA}\footnote{We use \texttt{MESA} version 24.08.1. All inlists and relevant files are available on Zenodo at \textit{The folder will be uploaded at the time of publication}};][and references therein]{Jermyn2023} from the MS to the WD cooling sequence to provide the background stellar structure for the magnetic field evolution calculation. To constrain the magnetic field, we choose the closest RG model to a typical oscillation mode frequency of $150\uHz$. At this evolutionary stage, we normalize our field configurations to match the signatures measured in \citet{Hatt2024}. To do so, we use the code \texttt{magsplitpy} and follow the method outlined in App.~\ref{app:kernel_avg} based on the red giant literature \citep{Das2020, Bugnet2021, Mathis2021, Li2022, Mathis2023, Bhattacharya2024, Das2024}. Our normalization then results in core-averaged squared radial field strengths $\langle B_r^2 \rangle$ that would be typically measured from observing oscillation modes for the chosen RG model (see App.~\ref{app:kernel_avg}). We take $\sqrt{\langle B_r^2 \rangle}=100$kG in Fig.~\ref{fig:intro}. Typical detected field strengths range between $\sim$10kG up to $\sim$200kG \citep{Hatt2024},  with the highest detection being above $\sim$600kG \citep{Deheuvels2023}.

\subsection{Numerical methods for the evolution of magnetic fields along stellar evolution}
We evolve magnetic fields via flux conservation and diffusion. Most work on WDs regarding this has relied on calculating ohmic diffusion time scales and did not consider the configuration of the magnetic field itself \citep[e.g.][C24]{Cumming2002}. However, more recent work has relied on the evolving magnetic fields on the WD cooling sequence from initial step function-like field configurations \citep[][CT26]{Castro-Tapia2024a, Blatman2025}. This approximation works well during the WD cooling phase, where the structural evolution of the star can be neglected. However, during earlier evolutionary stages, nuclear burning creates structural changes in the stellar interior, which can drastically change the size and radial distribution of the magnetized mass. Especially strong density gradients, such as around the hydrogen-burning shell (H-shell), can have a strong effect on the magnetic flux distribution. Thus, for earlier evolutionary stages, we need to take the changes in the stellar structure into account and instead solve the full induction equation:
\begin{equation} \label{eq:induction}
    \dt{\Vec{B}} = \curl \left(\Vec{u}\wedge\Vec{B}\right)-\curl\left(\eta\curl\Vec{B}\right)~,
\end{equation}
with $\Vec{B}$ the magnetic field, $\Vec{u}$ the fluid velocity, and $\eta$ the magnetic diffusivity. To solve it along the evolution, we follow the method described in \citet[][sec.~2.2]{Takahashi2021}. Using \texttt{MESA}, we calculate all relevant structure parameters. In particular, for the magnetic diffusivity, we follow the prescriptions of \cite{Potekhin1999} and \cite{Stygar2002} for the degenerate and non-degenerate regions, respectively, and linearly interpolate between the two formulations in the intermediate region. We neglect the small effect the crystallization of the WD core has on the magnetic diffusivity. For more details, see App.~\ref{app:mag_evo}.

\section{Results and Discussions}
\label{sec:results}
\subsection{Magnetic evolution before the white dwarf stage}
Figure~\ref{fig:intro} shows how fields from the different scenarios evolve alongside stellar evolution. We show that magnetic fields confined within the MS convective core (resulting from scenarios A and B) are buried deep below the H-shell during the red giant branch (RGB). The sensitivity of the modes to the magnetic field is dominant at this H-shell (see App.~\ref{app:kernel_avg}). Therefore, to be detected as a core-averaged 100kG field via asteroseismology on the RGB, it has to reach a huge central strength of the order of $\sim$10MG, which is about half of the critical field above which the mode core energy would be suppressed \citep{Fuller2015}.

However, Scenario C depicts a different picture, as the magnetic field extends beyond the H-shell. An interesting feature can be seen in the initial WD field for such Scenario C. Rather than being centered at the core, this field configuration peaks in a shell instead. This shell configuration comes from the steep density gradient around the H-shell during the RGB, which compresses the magnetic field as it slowly moves outwards. This only happens to the magnetic field of Scenario C because the other two fields vanish before reaching the H-shell (details on this effect are given in App.~\ref{app:whybump}).

\subsection{Emergence of the fossil magnetic fields at the surface}
\begin{figure}[ht]
    \centering
    \includegraphics[width=0.98\linewidth]{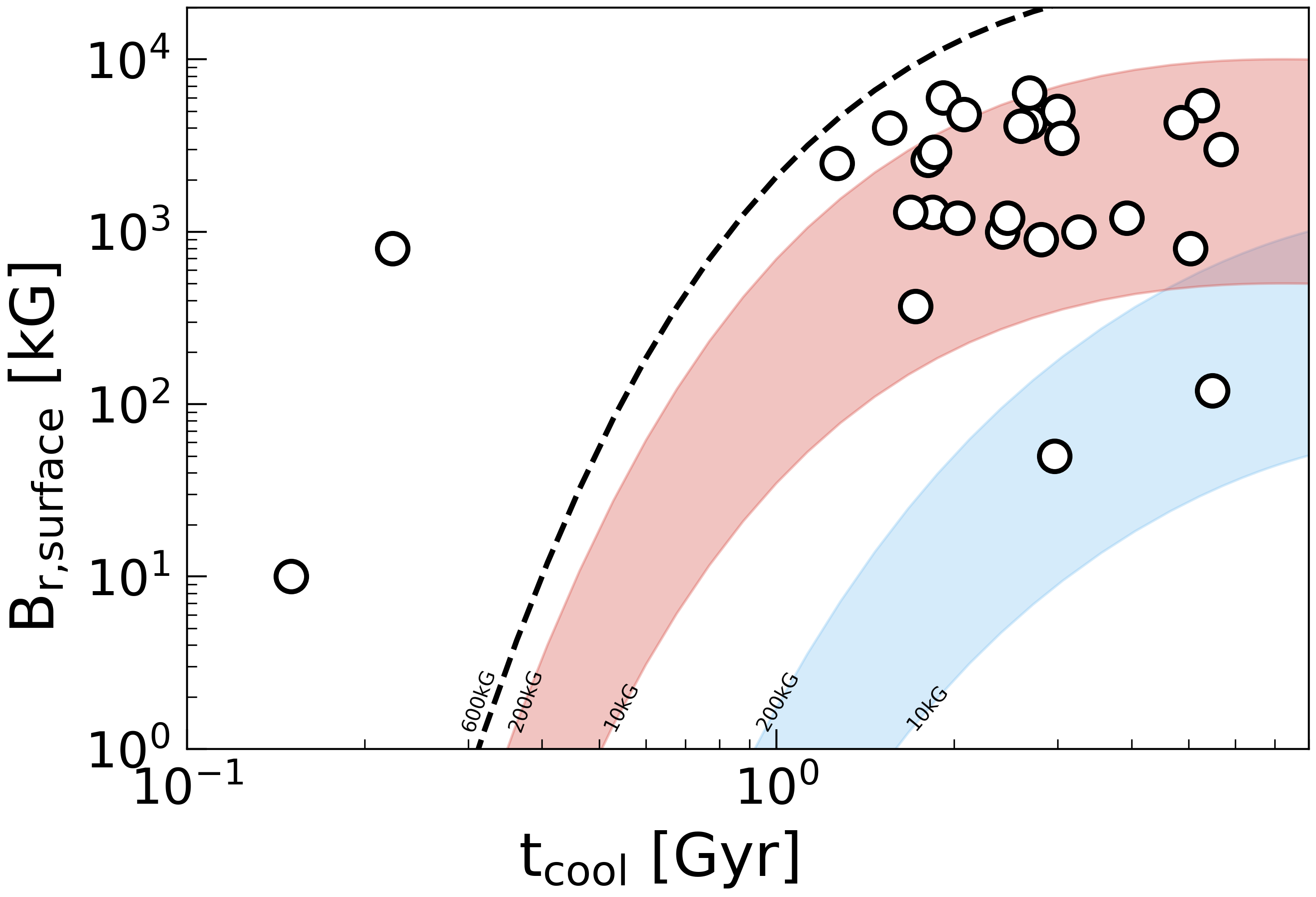}
    \caption{Radial surface field strength as a function of cooling age of the WD for scenarios B (blue) and C (red). The shaded regions show the evolution for varying field strengths on the RGB between $10$ (bottom) and $200$ (top) kG, as the range detected in \cite{Hatt2024}. The dashed line shows the emergence of the strongest detected RGB field of $600$kG \citep{Deheuvels2023} and places a lower limit on the FFS emergence timescale.
    The white circles show detected magnetic WDs from the samples of \citet{Bagnulo2022} and \citet{Moss2025} in the mass range $[0.5,0.64]\Msun$.}
    \label{fig:breakout_times}
\end{figure}

In Fig.~\ref{fig:breakout_times}, we summarize our results, calibrating when our various FFS result in the emergence of magnetic fields at the surface of WDs (see also App.~\ref{app:bump_diff}). We show that most detected magnetic WDs from \citet{Bagnulo2022} and \citet{Moss2025} can be explained by Scenario C with the core-averaged RGB field strengths detected in \cite{Hatt2024}, which range from 10kG to 200kG (red-shaded region). While there are also a few magnetic WDs outside this region, most of them can be explained by the higher RGB field strengths detected in \citet{Deheuvels2023}, which can reach at least 600kG (black-dashed line). Thus, almost all magnetic WDs in the mass range ($0.5\Msun$ to $0.64\Msun$) compatible with the mass distribution of detected magnetic RGs can be explained by a magnetic field that has a significant amplitude far above the H-shell during the RGB (Scenario C, see also App.~\ref{app:thresh} for more details).

In contrast, the magnetic field of Scenario B can only explain very few weakly magnetized old WDs. Even when normalizing the field to 600kG during the RGB, it can explain at most a third of the magnetic WDs shown in Fig.~\ref{fig:breakout_times}. For ease of reading, we choose not to show Scenario A in Fig.~\ref{fig:breakout_times} as its field strength is even lower than that of Scenario B. Two very young magnetic WDs can very clearly not be reached with any of our magnetic field configurations, and are incompatible with our formalism. 

To show the robustness of these conclusions, we discuss their dependence on stellar mass and on the extent of the magnetic field during the RGB in App.~\ref{app:robustness}.

\subsection{Conclusions on the fossil field scenario}
From Fig.~\ref{fig:breakout_times}, we conclude that fossil fields resulting from measured RG internal fields near the H-shell emerge at the surface of WDs in a timescale compatible with spectroscopic observations. We therefore confirm the result of C24 and CT26 that the FFS is a great candidate to explain WD magnetic field amplitudes and emergence timescales.

However, our study points out that, to link the magnetic fields detected in RG's cores to the WD stage, magnetic fields cannot result solely from convective core dynamo during the MS. This conclusion differs from that of C24 and CT26 due to the stable initial radial profile of the field we are considering following \citet{Braithwaite2008, Broderick2007,  Duez2010} as opposed to a constant radial magnetic energy in C24 and CT26.
Such an extended strong field at the end of the MS is not realistic if considering only the stabilization of the core dynamo effect, during which the reconnection of the field results in a loss of magnetic energy,  with a flattened radial energy distribution decaying much faster compared to more tapered distributions (this work). Thus, even though the core dynamo on the MS can create fields on the order of $\sim1-100$kG, constant radial field distributions are likely to stabilize into much weaker fields compared to more realistic profiles once the convection ends.
In addition, the resulting stable large-scale poloidal component of the field \citep{Braithwaite2008} is likely confined in the previously magnetized region \citep{Broderick2007, Duez2010}.

 We also emphasize the recent inference of a nearly pure toroidal field at the outer boundary of the convective core during the MS \citep{Takata2026}, which reinforces that, if rotational effects are not at play, the radial component of the magnetic field remains contained in the convective core \citep[also in agreement with massive star convective core simulations, e.g.][]{Ratnasingam2024}).
 The strong uniform field used in CT26 is in fact more representative of an extended magnetized radiative zone as our Scenario C, which could result from the convective core dynamo coupled with a fossil field in the MS radiative zone \citep{Hidalgo2025}, from the coupled effects of core dynamo and of fast rotation during the MS \citep[][enhancing the advection of the core dynamo in the outter layers]{Hidalgo2024}, or directly from a dynamo action in RGs radiative interiors \citep[e.g.][therefore unlinked to MS core dynamo]{Fuller2019}.

Scenario C leads to a radial magnetic field profile confined in a shell around $0.35$ of the stellar radius at the start of the WD cooling sequence. This structural effect is due to the local magnetic flux conservation in its mass coordinate, compressing the field around the hydrogen shell burning layers during the RGB (see Appendix~\ref{app:whybump} for the detailed origin of the confinement). The long-term stability of such a geometry, which differs from the Prendergast-like stable profiles as in \citet{Kaufman2022}, remains to be studied (this radial magnetic profile is, however, robust against magnetic diffusion on Giga-year timescales, see Appendix~\ref{app:bump_diff}). A second possible extension of our work is to include the impact of the formation of the small convective core during the core-helium-burning phase, as it is expected to generate additional magnetic energy in the core, potentially strengthening the deep radial magnetic profile even further \citep[see similar configuration during the MS in][]{Featherstone2009}. Both analyses require dedicated 3D simulations and are out of the scope of our study.

Under the FFS, stars sustain their magnetic fields all along their evolution. Distinguishing between the FFS and the crystallization scenario, for example, from \cite{Isern2017} or contemporary fields in radiative zones, as in \cite{Fuller2019}, is crucial for a better constraint on angular momentum and chemical transport in stars. The new area of magnetoasteroseismology on the RGB, MS, and even young WD \citep[e.g.][]{Rui2025} is key to further investigating the possibility of magnetic survival across stellar ages.

\begin{acknowledgements}
    The authors thank the referee for their helpful and constructive report, which has significantly enhanced the quality of the manuscript. The authors thank I. Caiazzo, L. Ferrario, and L. Buchele for very useful discussions. L. Barrault, L. Bugnet, and L. Einramhof gratefully acknowledge support from the European Research Council (ERC) under the Horizon Europe programme (Calcifer; Starting Grant agreement N$^\circ$101165631). L. Barrault acknowledges the support of the Austrian Academy of Sciences through the Doctoral Fellowship Programme (DOC) of the Austrian Academy of Sciences 27648. While partially funded by the European Union, views and opinions expressed are, however, those of the authors only and do not necessarily reflect those of the European Union or the European Research Council. Neither the European Union nor the granting authority can be held responsible for them.
\end{acknowledgements}

\bibliographystyle{bibtex/aa}
\bibliography{refs}

\appendix

\section{Magnetic evolution scenarios} \label{app:mag_scenarios}
In this work, we consider three different fossil field scenario (FFS) candidates to produce the magnetic fields at the surface of WDs, guided by various formation mechanisms. 

\subsection{Scenario A: A Convective core dynamo during the main sequence}
\label{app:scenario_a}
We first assume that the magnetic field is generated during the MS, in the convective cores of A/B type stars by convective dynamo action. This field then is expected to relax into a stable dipolar configuration, according to, for instance, the simulation by \citep{Braithwaite2008}, at the end of the MS when the core becomes radiative. Thus, we assume a stable dipole configuration as defined in App.~\ref{app:mag_geom} that is confined in the mass coordinate of the maximum extent of the MS convective core (Scenario A; see the darkblue shell in Fig.~\ref{fig:intro} and App.~\ref{app:mag_geom} for the field geometry). More specifically, we define the mass of the magnetized core $\mathcal{M}_A$ as the maximum mass coordinate during the MS that has been part of the convective core. For the $1.5\Msun$ star modeled in this paper, we have $\mathcal{M}_A = 0.13\Msun$.

\subsection{Scenario B: An Extended convective dynamo during the main sequence}
Alternatively, we relax the strict condition of confinement in $\mathcal{M}_A$. Indeed, magnetic diffusion during the MS could extend the field lines beyond the convective mass, and various overshooting prescriptions in the model generate uncertainties on the extent of the convective core. To extend the magnetized mass, we allow in Scenario B for a strong diffusion of the dynamo field into the outer radiative zone during the MS (see the lightblue shell in Fig.~\ref{fig:intro}). For this, we use the ohmic diffusion relation
\begin{equation} \label{eq:ohmic_diffusion}
    \tau_{\Omega} = \frac{L_{\Omega}^2}{\eta} \iff L_\Omega = \sqrt{\tau_\Omega \eta}\, ,
\end{equation}
where $L_{\Omega}$ is the distance the magnetic field diffuses in the time $\tau_{\Omega}$ given the magnetic diffusivity $\eta$. In our case we have $L_\text{diff} = \sqrt{t_{\text{MS}}\eta_{\text{MS}}}$, which we transform into mass coordinates $\mathcal{M}_\text{diff}$ as additional magnetic mass. Furthermore, we multiply $\mathcal{M}_\text{diff}$ by three to give an upper bound on the effect of this additional mass due to the above-mentioned processes. We calculate the convective core mass as in subsection~\ref{app:scenario_a} and then add the additional mass coordinate into which the magnetic field has diffused. This gives a total magnetic mass of
\begin{equation}
    \mathcal{M}_B = \mathcal{M}_A + 3\mathcal{M}_\text{diff}.
\end{equation}
For the $1.5\Msun$ star modeled in this paper, we have $\mathcal{M}_B = 0.15\Msun$, which does not significantly differ from $\mathcal {M}_A$. Alternatively to very efficient magnetic diffusion, one could also view this additional mass as the uncertainty in the size of the convective core during the main-sequence. For example, adding overshooting into the stellar evolution models can also increase the size of the convective core by $\sim20\%$ \citep{Paxton2013}, which is similar to the chosen mass increase for this scenario.

\subsection{Scenario C: A Magnetized radiative interior during the red giant phase}
Lastly, we assume that the stable magnetic field configuration fills the full radiative interior during the RGB when the star reaches a typical oscillation mode frequency of $\numax = 150\uHz$ (Scenario C; see the red shell in Fig.~\ref{fig:intro}). This is the magnetic field extent used by the red giant community when prescribing asteroseismic magnetic signatures \citep[e.g.][]{Bugnet2021}. As can be seen in Fig.~\ref{fig:intro}, we cannot only rely on the convective core dynamo during the MS alone to create this field. Some additional process is needed such that there is also a magnetic field present in parts of the radiative envelope during the MS. For example, fast rotation might leak the MS convective core field into the radiative envelope along the rotation axis of the star \citep{Hidalgo2024}. More importantly, there could already be some large-scale stable magnetic field present in the radiative envelope during the MS \citep{Hidalgo2025}, which was created during stellar formation. Other possibilities are the Tayler-Spruit dynamo and assimilates \citep[e.g.][]{Spruit2002, Fuller2019} which might play a role in generating the magnetic field directly in the radiative interior during the RGB. Scenario C  leads to a magnetic mass of $\mathcal{M}_C = 0.26\Msun$ on the early RGB ($\numax = 150\mu$Hz).

\section{Magnetic configurations} \label{app:mag_geom}

\begin{figure}[b]
    \centering
    \includegraphics[width=0.98\linewidth]{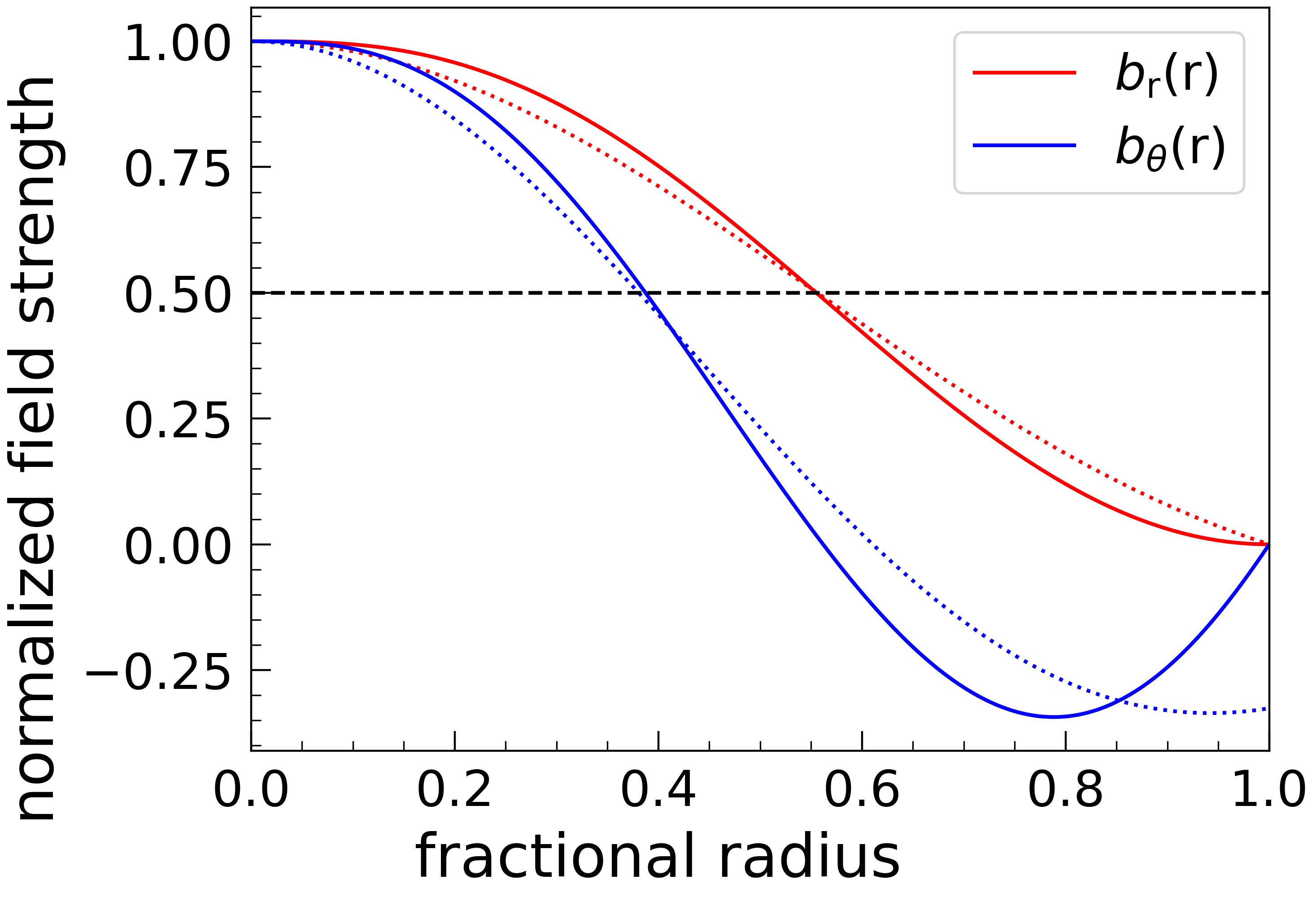}
    \caption{Comparison between $b_r(r)$ and $b_\theta(r)$ (Eqs.~\ref{eq:br_def}-\ref{eq:bt_def}) from \citet{Broderick2007} (dotted lines) and the polynomial fit that enforces the correct boundary conditions (full lines) as a function of the fractional radius of the magnetized zone. The fit is created such that both formalisms agree at half the strength of $b_r(r)$ (black dashed line).}
    \label{fig:bn_vs_fit}
\end{figure}

In this paper, we consider poloidal, axisymmetric, dipolar field configurations. These choices are supported by the following. 
First, ignoring the turbulent mean field dynamo, the poloidal and toroidal components evolve uncoupled. As the toroidal part vanishes at the surface, it does not contribute to the observed WD fields. Thus, we ignore that component and focus on the evolution of the poloidal component.
Second, \citet{Braithwaite2008} demonstrated that stable non-axisymmetric fields are significantly weaker than axisymmetric ones and less stable.
Finally, in addition to the dipolar component being the dominant term in the expansion of the magnetic field, \citet{Hardy2023} also showed that most magnetic WDs are well fitted with a dipolar configuration. Therefore, the full expression of our magnetic field is written as follows:
\begin{align}
    \Vec{B} &= \Vec{B_r} + \Vec{B_\theta}\, , \\
    \Vec{B_r} &= B_0b_r(r)\Vec{\hat{Y}}_{10}(\theta,\phi)\, , \label{eq:br_def}\\
    \Vec{B_\theta} &= B_0b_\theta(r)\Vec{\hat{\Psi}}_{10}(\theta,\phi)\, , \label{eq:bt_def}
\end{align}
with the vector spherical harmonics $\Vec{\hat{Y}}_{\ell,m} = \Vec{\hat{e}}_rY_{\ell,m}$ and $\Vec{\hat{\Psi}}_{\ell,m} = r\Vec{\nabla} Y_{\ell,m}$ are defined as in \citet{Barrera1985}. The $Y_{\ell,m}$ are the usual spherical harmonics.

For the radial profiles $b_r(r), b_\theta(r)$, we created polynomial fits to the stable profiles in \citet{Broderick2007}, as they do not fulfill the boundary conditions discussed in App.~\ref{app:mag_evo} because their assumptions force $b_\theta(r)$ to be discontinuous at the magnetic boundary. Thus, we fit a polynomial to these profiles while enforcing the correct boundary conditions (see Fig.~\ref{fig:bn_vs_fit}). This polynomial is parametrized by $\rthresh$, the fractional radius of the magnetized zone at which $b_r(r)$ reaches half strength compared to the center. The best fit to the profiles from \citet{Broderick2007} is achieved with $\rthresh = 0.555$, which is the value used for all fields in the main text.

\section{Constraints from asteroseismology during the RGB} \label{app:kernel_avg}
We use the recent asteroseismic measurements of the magnetic field strength in the radiative interior during the RGB to constrain the initial field strength at the start of the WD cooling sequence. These measurements are mostly sensitive to the radial field component averaged at the H-shell, with some additional sensitivity closer to the core \citep{Li2022, Bhattacharya2024}. Following App.~C in \citet{Bhattacharya2024}, we can define the measured squared field strength as
\begin{equation}
    \langle B_r^2\rangle = \frac{1}{4\pi}\frac{\int_0^{R_c}K(r)\left(\iint\Vec{B_r}\cdot\Vec{B_r}\sin\theta d\theta d\varphi\right) r^2dr}{\int_0^{R_c}K(r)r^2dr}\, ,
\end{equation}
which is a horizontal and radial average of the magnetic field weighted by the sensitivity kernel $K(r)$ \citep[for details see][]{Das2020, Bhattacharya2024}, with $R_c$ the core radius. Thus, we use detected values for $\sqrt{\langle B_r^2\rangle}$ from \citet{Deheuvels2023} and \citet{Hatt2024} to calibrate $B_0$ (Eqs.~\ref{eq:br_def}-~\ref{eq:bt_def} and Fig.~\ref{fig:bn_vs_fit}).

\section{Details on the magnetic evolution} \label{app:mag_evo}
We describe in detail in this appendix the method we use to evolve magnetic fields along stellar evolution. Most of this appendix is a direct translation of the method described in \citet{Takahashi2021}.

\subsection{Equations}
First, we assume that the magnetic fields we consider are weak enough not to affect the stellar structure. Thus, we can decouple the evolution of the magnetic field from that of the considered star. Additionally, for simplicity, we ignore all effects from the mean-field MHD-dynamo formulation, which results in the ohmic induction equation (also Eq.~\ref{eq:induction}):
\begin{equation}\label{eq:app_induction}
    \dt{\Vec{B}} = \curl \left(\Vec{u}\wedge\Vec{B}\right)-\curl\left(\eta\curl\Vec{B}\right).
\end{equation}
In the case of zero diffusivity, magnetic flux will be conserved in the originally magnetized mass coordinate. If there is finite diffusivity, instead, the magnetic flux evolves according to Alfven's theorem. Given a surface S that moves with velocity $\Vec{U}$, the magnetic flux $\Phi_B = \int_S\Vec{B}\cdot d\Vec{S}$ evolves according to
\begin{equation}\label{eq:alfven}
    \frac{d\Phi_B}{dt} = \int_S\left(\dt{\Vec{B}} - \curl\left(\Vec{U}\wedge\Vec{B}\right)\right)\cdot d\Vec{S}.
\end{equation}

We consider the same surface $S$ as \citet{Takahashi2021}, which is a polar cap at radius $r_c$ from the center with an opening angle $\theta_c$ (see \citet{Takahashi2021} Fig.~1 for an illustration of $S$). Additionally, we define $\Vec{U}$ as the radial velocity the stellar model has at $r_c$, i.e., $\Vec{U}(t)~=~\Vec{u}(r_c,t)$. Thus, inserting Eq.~\ref{eq:app_induction} into Eq.~\ref{eq:alfven} and using Stokes' Theorem results in 
\begin{equation}\label{eq:flux_evo}
    \frac{d\Phi_B}{dt} = -\oint_{\partial S}\bigg(\eta\curl\Vec{B}\bigg) \cdot d\Vec{l}\, ,
\end{equation}
where $\partial S$ is the boundary of S. Since the magnetic field $\Vec{B}$ has zero divergence, i.e., $\Vec{\nabla}\cdot\Vec{B}=0$, we can define a vector potential $\Vec{A}$ such that $\Vec{B} = \curl\Vec{A}$. Since we are only interested in a poloidal magnetic field, we can define $\Vec{A}$ as
\begin{equation} \label{eq:vp}
    \Vec{A} = B_0\frac{X(r)}{r}\Vec{\Upsilon}_{\ell,m}\, ,
\end{equation}
with $\Vec{\Upsilon}_{\ell,m} = \Vec{r}\wedge\Vec{\nabla}Y_{\ell,m}$ the toroidal vector spherical harmonics\footnote{Note that we use $\Vec{\Upsilon}$ here instead of $\Vec{\Phi}$ used in \citet{Barrera1985} to avoid confusion with the magnetic flux.} \citep{Barrera1985}, $B_0$ the field normalization from Eq.~\ref{eq:br_def}, and $X$ such that
\begin{align}
    b_r(r) &= -\frac{\ell(\ell+1)}{r^2}X(r), \\
    b_\theta(r) &= -\frac{1}{r}\dr{X}(r).
\end{align}
The relation between $b_r, b_\theta$ and $X$ follows directly from taking the curl of Eq.~\ref{eq:vp} and comparing terms with Eqs.~\ref{eq:br_def}-\ref{eq:bt_def}.

Making use of the vector spherical harmonics curl identities \citep{Barrera1985} we can write the right hand side of Eq.~\ref{eq:flux_evo} as
\begin{equation}
    -\frac{B_0}{r_c}\eta\bigg(\frac{\ell(\ell+1)}{r_c^2}X(r_c) - X''(r_c)\bigg)\oint_{\partial S}\Vec{\Upsilon}_{\ell,m}\cdot d\Vec{l}
\end{equation}
where $f'= \partial f/\partial r$, and the magnetic flux as
\begin{align}
    \Phi_B(r_c,\theta_c,t) &= \int_S\bigg(\curl\Vec{A}\bigg)\cdot d\Vec{S} = \oint_{\partial S}\Vec{A}\cdot d\Vec{l}, \\
    &= B_0 \frac{X(r_c,t)}{r_c}\oint_{\partial S}\Vec{\Upsilon}_{\ell,m}\cdot d\Vec{l}\, .
\end{align}
Notice that the only place $r_c$ comes into play in the integral is in $\partial S$ as the distance of the surface from the center. Thus, we can rescale $S$ to take out this dependence. This results in
\begin{equation}
    \oint_{\partial S}\Vec{\Upsilon}_{\ell,m}\cdot d\Vec{l} = r_c\oint_{\partial \bar{S}}\Vec{\Upsilon}_{\ell,m}\cdot d\bar{\Vec{l}} = r_c \bar{\mathcal{F}}(\theta_c)
\end{equation}
where quantities denoted as $\Bar{\square}$ are now scaled to be a unit distance away from the center. Thus, we can write the integral over the boundary of $S$ as $r_c$ times a function that only depends on the opening angle of $S$. Thus, $\bar{\mathcal{F}}$ does not depend on time and can be pulled out of the total time derivative of $\Phi_B$. Using this, we can see that $X$ is just the magnetic flux scaled by a constant that just depends on the surface S through which the magnetic flux is defined, i.e.,
\begin{equation}\label{eq:flux_equivalence}
    \Phi_B(r_c,\theta_c) = X(r_c) B_0 \bar{\mathcal{F}}(\theta_c)\, .
\end{equation}

Finally, combining everything, we get
\begin{equation}\label{eq:flux_diffusion}
    \frac{dX}{dt}(r,t) = \eta\left(X''(r) - \frac{\ell(\ell+1)}{r^2}X(r,t)\right)
\end{equation}
with $\square '$ and $\square''$ the first and second spatial derivatives respectively, and canceled $B_0\bar{\mathcal{F}}$ on both sides. Since $r_c$ is arbitrary, we leave out the subscript $c$ and end up with the general evolution equation for the magnetic flux. \\

One complexity we have not discussed so far is the implicit time-dependence of $r$, i.e., everywhere we should replace $r$ with $r(t)$. This is important to consider since the left-hand side of the equation is the total time derivative of $X$. We solve this by applying a transformation from radius to mass coordinates. We define mass coordinates $\Vec{m} = (m, \theta, \phi)$ with
\begin{equation}\label{eq:m_r_relation}
    m(r(t), t) = \int_0^{r(t)}4\pi s^2\rho(s,t)ds = \int_0^{r(t)}\rho_r(s,t)ds\, ,
\end{equation}
instead. Since the magnetic field is confined deeply in the star, we consider the magnetic flux to be conserved (as opposed to the work of \citet{Takahashi2021} where magnetic flux is lost by winds). Thus, we can define a maximum mass $\mathcal{M}$ that we consider for the magnetic evolution. This mass in turn defines a time dependent maximum radius $\mathcal{R}(t)$ via Eq.~\ref{eq:m_r_relation}:
\begin{equation}
    \mathcal{M} = \int_0^{\mathcal{R}(t)}\rho_r(s,t)ds\, .
\end{equation}
This way, our mass coordinate is independent of time, and we can discretize $\Vec{m}$ as a uniform grid on which to numerically solve Eq.~\ref{eq:induction}. We define functions in $\Vec{m}$ coordinates as
\begin{align}
    \Tilde{f}(\Vec{m}, t) = f(\Vec{r}(\Vec{m}, t), t)\, .
\end{align}
Additionally, to transform derivatives, we use
\begin{equation}\label{eq:basis_change}
    \partial_r = \frac{\partial m}{\partial r}\partial_m = \rho_r\partial_m.
\end{equation}
Thus, the final form of our magnetic evolution equation is
\begin{equation}
    \label{eq:final}\frac{d\Tilde{X}}{dt} = \Tilde{\eta}\left(\Tilde{\rho}_r\partial_m\left(\Tilde{\rho}_r\partial_m \Tilde{X}\right) - \frac{\ell(\ell+1)}{\Tilde{r}^2}\Tilde{X}\right).
\end{equation}

We numerically solve Eq.~\ref{eq:final} along the evolution of the model star, starting from the end of the MS, resulting in the field profiles represented in Fig.~\ref{fig:intro}.

\subsection{Boundary conditions}
We follow the mass coordinates enclosed in the radius $\mathcal{R}(t)$ comprising the final WD mass $\mathcal{M}$ during the full evolution, and follow the boundary conditions as in \citet{Cumming2002} for a dipolar field:
\begin{align}
    \dr{X}(0) &= \frac{2}{r}X(0)\, , \\
    \dr{X}(\mathcal{R}(t)) &= -\frac{1}{r}X(\mathcal{R}(t))\, ,
\end{align}
which forces the magnetic field to be finite at the center and continuous with a vacuum field at the outer boundary. We use the vacuum field boundary also for the $\mathcal{R}(t)$ boundary condition during the evolution pre WD to allow for diffusion of the magnetic field. We compared results obtained from this boundary condition to one that confines the flux in the considered zone, i.e., setting $X(\mathcal{R}(t))=0$. However, due to the magnetic fields being confined deep below the chosen boundary, the difference between these two outer boundary conditions is negligible.

To transform these equations into mass coordinates, we apply Eq.~\ref{eq:basis_change} to get
\begin{align}
    \dm{\Tilde{X}}(0) &= \frac{2}{\Tilde{r}\Tilde{\rho}_r}\Tilde{X}(0)\, , \\
    \dm{\Tilde{X}}(\mathcal{M}) &= -\frac{1}{\Tilde{r}\Tilde{\rho}_r}\Tilde{X}(\mathcal{M})\, .
\end{align}

\section{Evolution of the magnetic field geometry during the RGB}\label{app:whybump}
\begin{figure}[t]
    \centering
    \includegraphics[width=1\linewidth]{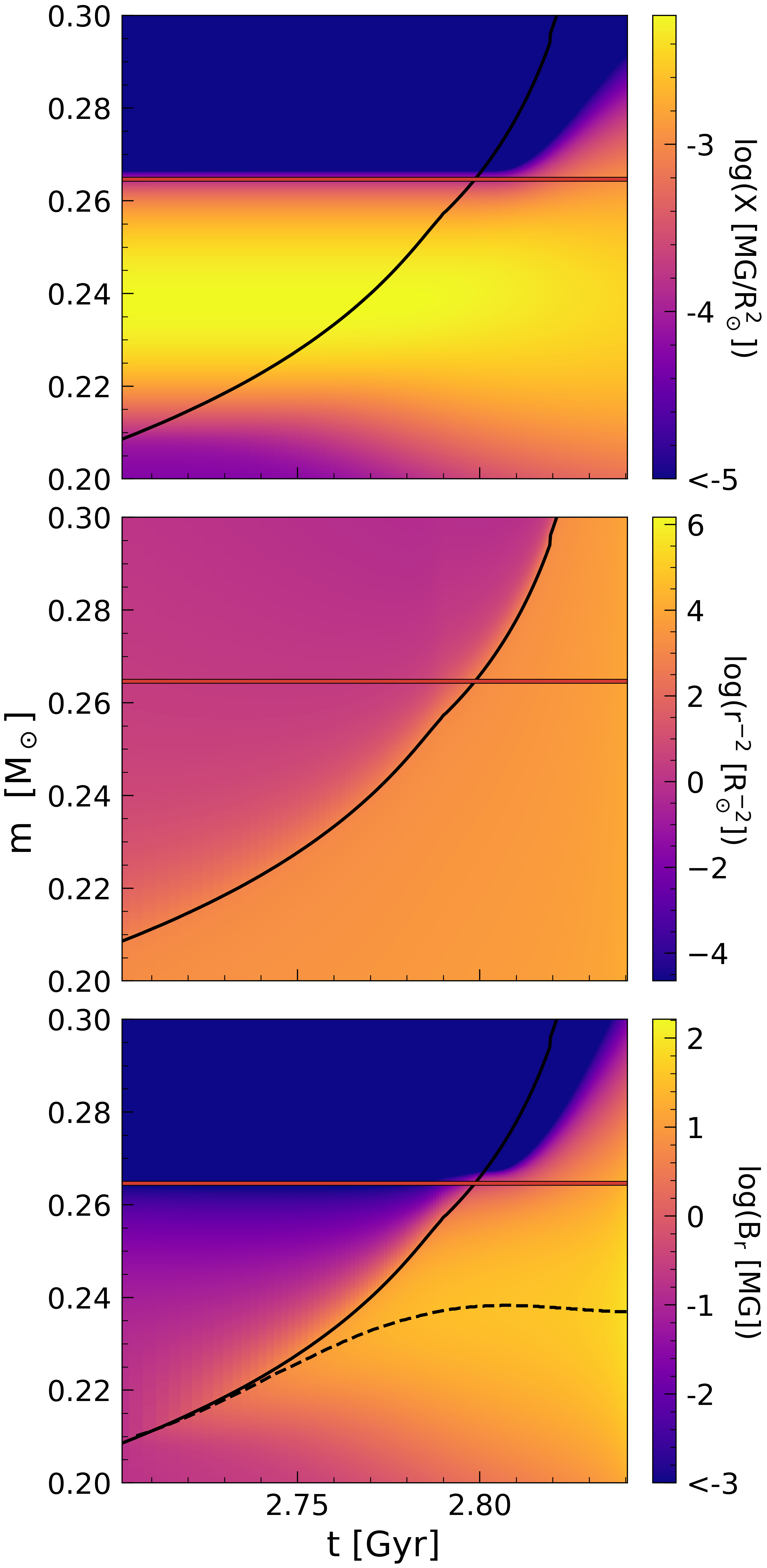}
    \caption{Evolution of Scenario C during the RGB around the hydrogen-burning shell (solid black line). The magnetic mass $\mathcal{M}_C\sim0.26\Msun$ is shown as a solid red line. Left: Contour map of the magnetic flux in log-scale as it evolves according to Eq.~\ref{eq:final}. Middle: Contour map of $r^{-2}$ in log-scale. Right: Contour map of $b_r$ in log-scale as a result of multiplying the two panels above. The dashed black line shows the location of the maximum of $b_r$.}
    \label{fig:whybump}
\end{figure}

\begin{figure}[ht]
    \centering
    \includegraphics[width=0.9\linewidth]{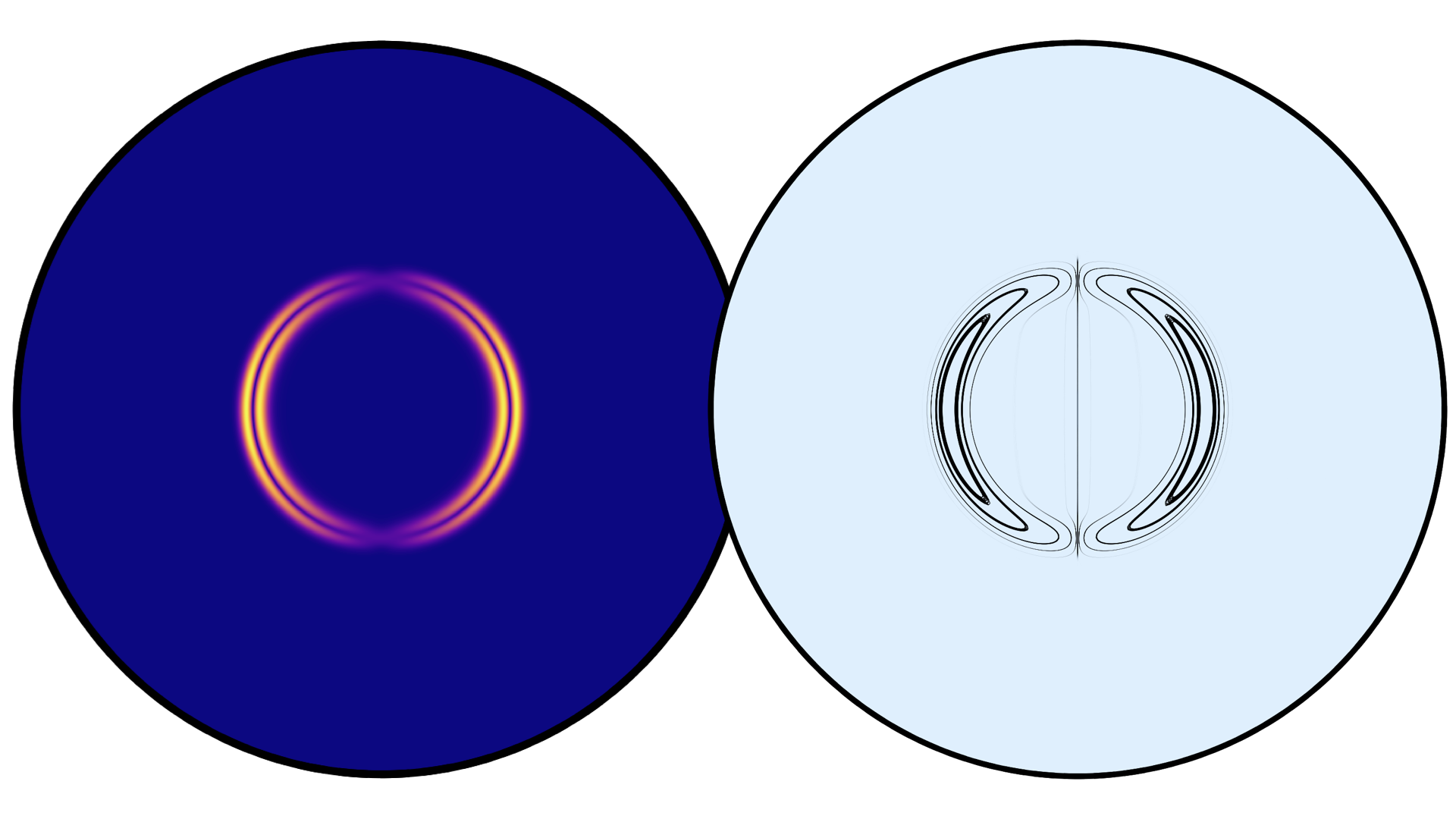}
    \caption{A 2D slice of the magnetic field of Scenario C at the start of the WD cooling sequence. Left: The local magnetic field strength $\sqrt{\Vec{B}\cdot\Vec{B}}$ throughout the full WD. The yellow region shows where most of the magnetic field strength is located. Right: The corresponding field lines to the same magnetic field shown on the left.}
    \label{fig:2d_field_slice}
\end{figure}

In this appendix, we provide insight into the mechanism driving the evolution of the magnetic field geometry of Scenario C, i.e., why the magnetic field in Scenario C becomes confined in a shell away from the center. 
First, we note that $X$ is just magnetic flux scaled by some constant that only depends on the choice of the surface through which the flux is defined (see Eq.~\ref{eq:flux_equivalence}). The magnetic flux is bound to its mass coordinate and only evolves via diffusion (see the left panel of Fig.~\ref{fig:whybump}). However, to get to $b_r$ we need to multiply $X$ by $r^{-2}$ (see the middle panel of Fig.~\ref{fig:whybump}), which decreases sharply above the H-shell (full black line) due to the strong density gradient present there. Thus, the resulting radial magnetic field strength $b_r$ reaches its maximum close to the flux maximum ($\sim0.24\Msun$ in this case; see the right panel of Fig.~\ref{fig:whybump}). A visualization of the magnetic field geometry is represented in Fig.~\ref{fig:2d_field_slice}.

In contrast, for the other scenarios, the flux maximum is confined below the hydrogen-burning shell, and thus, the strong density gradient there does not affect the magnetic field significantly.

\section{Magnetic evolution during the WD cooling sequence}\label{app:bump_diff}
\begin{figure}[b]
    \centering
    \includegraphics[width=1\linewidth]{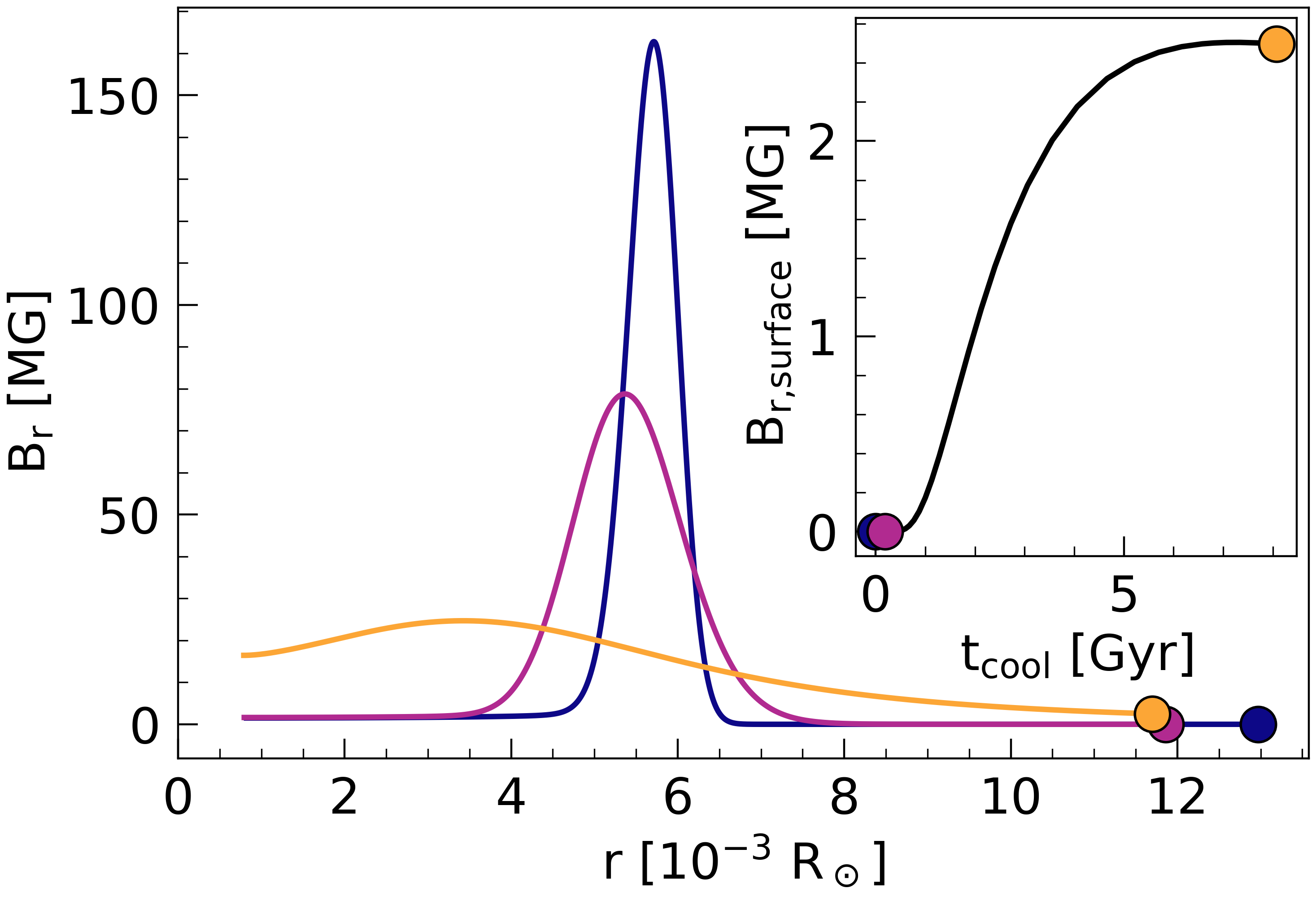}
    \caption{The evolution of Scenario C along the WD cooling sequence. Left: $B_r$ as a function of radius for three different cooling ages: $\tcool=0$Gyr (darkblue), $\tcool=0.2$Gyr (purple), $\tcool=8$Gyr (yellow), with the colored circles emphasizing the current field strength at the surface. Right: $B_r$ at the surface of the WD as a function of cooling age. The three colored circles correspond to the surface field strength of the three radial profiles in the left panel.}
    \label{fig:bump_evo}
\end{figure}

Here, we discuss the evolution of the radial dependence of the magnetic field along the WD cooling sequence. Fig.~\ref{fig:bump_evo} shows $B_r$ of Scenario C for different cooling ages. Already after 0.2Gyr (purple profile), the shape of the magnetic field has changed a lot compared to the initial profile (blue). However, the surface field strength (Fig.~\ref{fig:bump_evo} right panel) has not increased noticeably by that point (see also Fig.~\ref{fig:breakout_times} for comparison).
Even after 8 Gyr, the maximum of the magnetic field is still located in a shell away from the center of the WD. This shows that the concentration of the magnetic field in a spherical shell is stable against magnetic diffusion. As a visual guide, Fig.~\ref{fig:2d_field_slice} shows a 2D slice of the magnetic field at the start of the WD cooling sequence.
\section{Robustness of the results} \label{app:robustness}
In this appendix, we present additional checks to confirm that an extended magnetic field is needed during the RGB to explain the observed field strengths of magnetic WDs. First, we discuss the effect of varying the parameter $\rthresh$ (as defined in App.~\ref{app:mag_geom}) and afterwards, the effect of stellar mass.

\subsection{Impact of varying $\rthresh$}\label{app:thresh}
\begin{figure}
    \centering
    \includegraphics[width=1\linewidth]{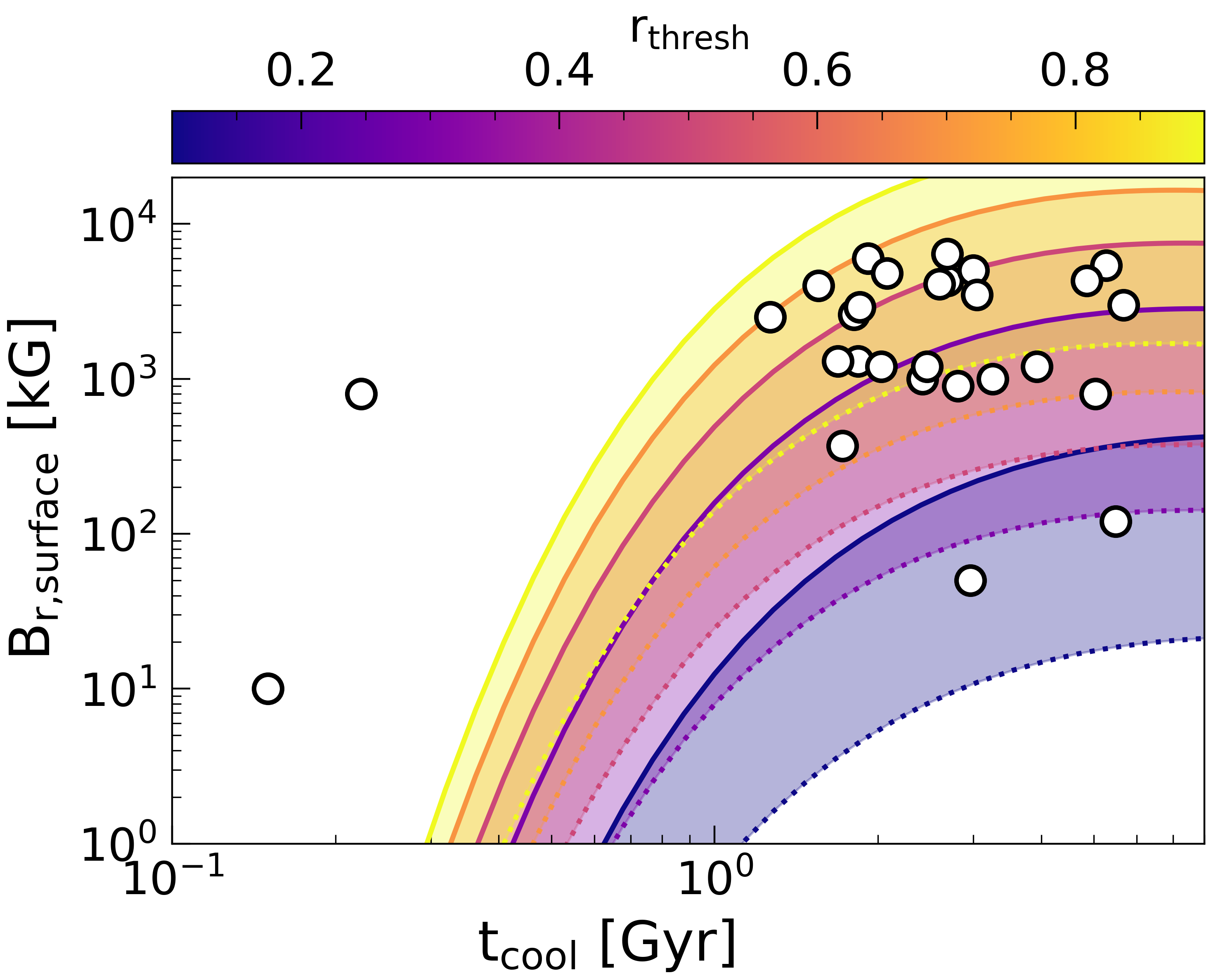}
    \caption{Same as scenario C in Fig. \ref{fig:breakout_times} but varying $\rthresh$.}
    \label{fig:breakout_diff_thresh}
\end{figure}

Here, we constrain the extent of the magnetic field needed during the RGB to fit the observed field strengths on the WD cooling sequence. For this, we modify scenario C by varying $\rthresh$ (see App.~\ref{app:mag_geom}). Fig.~\ref{fig:breakout_diff_thresh} shows that an $\rthresh \gtrsim 0.3$ is needed to explain most observed field strengths on the WD cooling sequence. Thus, we conclude that a significant portion of the radiative interior during the RGB has to be strongly magnetized, extending much beyond the H-shell.

\subsection{Impact of the stellar mass}\label{app:mass_effect}
\begin{figure}
    \centering
    \includegraphics[width=1\linewidth]{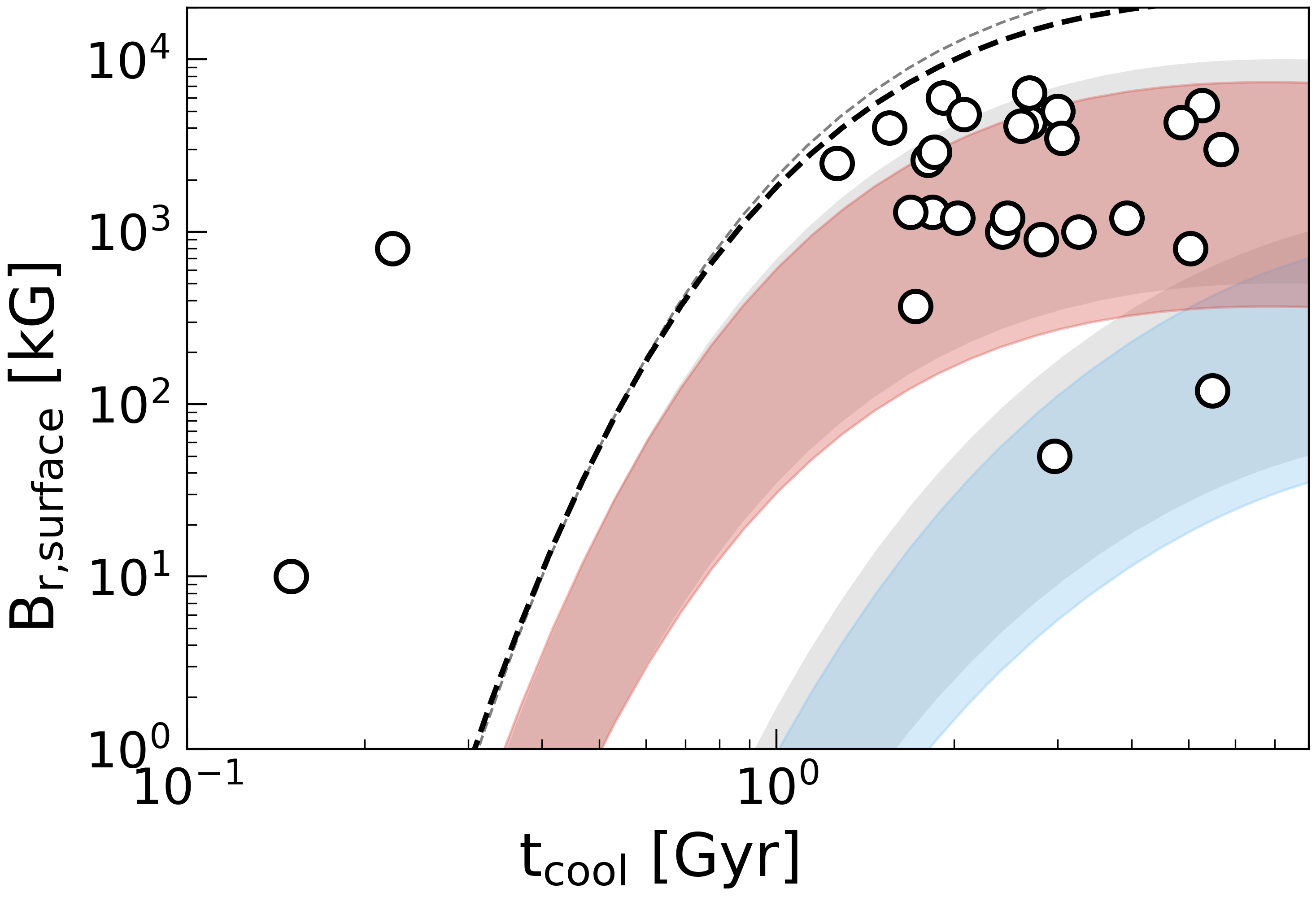}
    \caption{Same as Fig. \ref{fig:breakout_times} but for a stellar model with initial mass of $1.3\Msun$. The field evolutions for the $1.5\Msun$ model from Fig. \ref{fig:breakout_times} are shown in grey for easy comparison.}
    \label{fig:breakout_low}
\end{figure}

\begin{figure}
    \centering
    \includegraphics[width=1\linewidth]{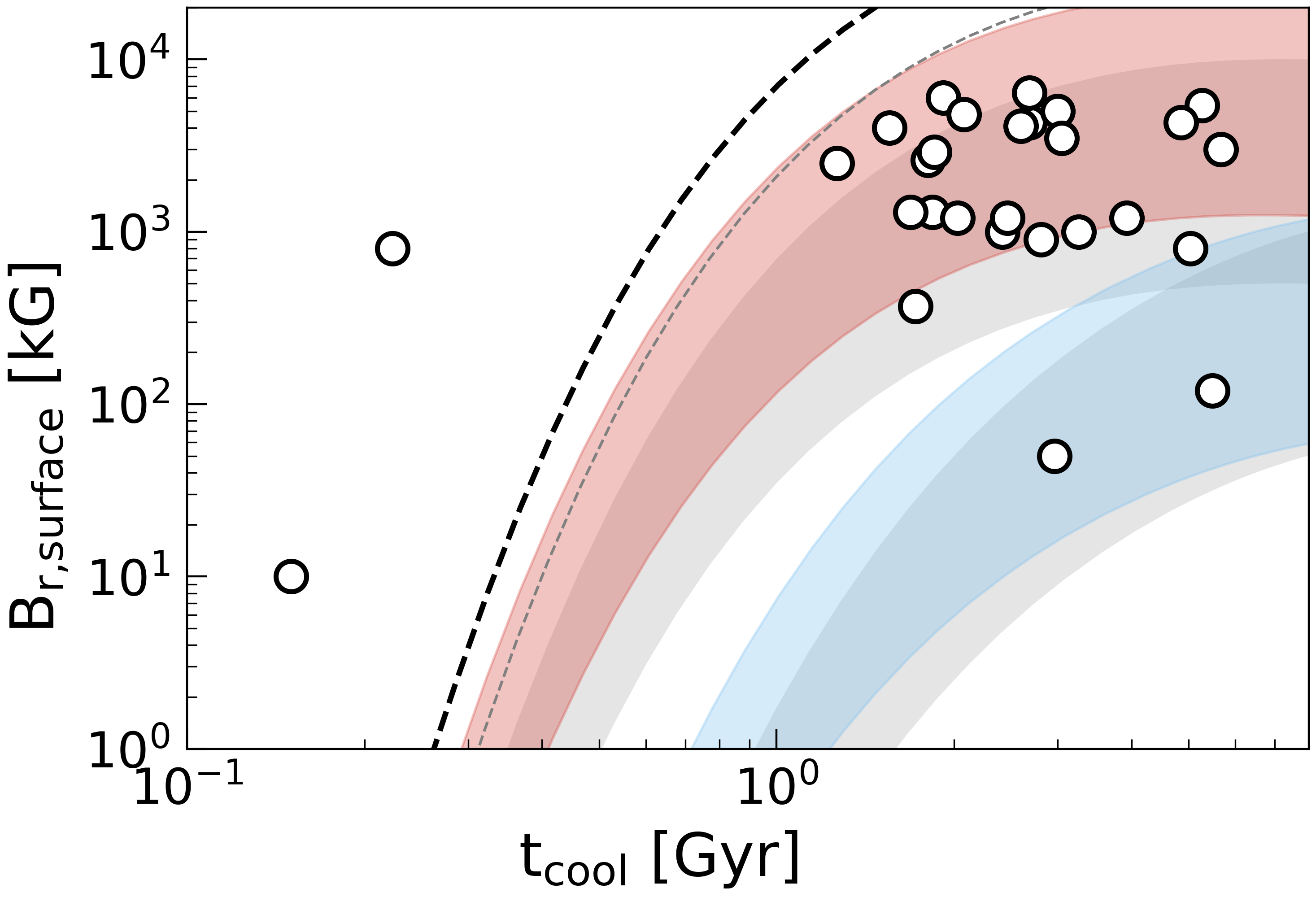}
    \caption{Same as Fig. \ref{fig:breakout_times} but for a stellar model with initial mass of $1.8\Msun$. The field evolutions for the $1.5\Msun$ model from Fig. \ref{fig:breakout_times} are shown in grey for easy comparison.}
    \label{fig:breakout_high}
\end{figure}

To characterize the mass dependence of the magnetic emergence time on WD surfaces, we show the same plot as Fig. \ref{fig:breakout_times}, but for initial masses of $1.3\Msun$ (Fig. \ref{fig:breakout_low}) and $1.8\Msun$ (Fig. \ref{fig:breakout_high}), representing typical lower and upper mass bound of magnetized RGs that used to have a convective core during the MS \citep{Hatt2024}. While there are small shifts in both the emergence time and the maximum field strength reached, these changes are not large enough to change our results as stated in Section~\ref{sec:results}. Thus, we validate our conclusion that an extended radial field profile is needed on the RGB to explain the observed WD surface field strengths from asteroseismic constraints on the RGB, independently of the initial mass of the star.

\end{document}